\documentclass[a4paper,12pt]{article}

\usepackage{moreverb,url, dirtytalk, authblk, caption, subcaption, graphicx}
\usepackage[colorlinks,bookmarksopen,bookmarksnumbered,citecolor=blue,urlcolor=blue]{hyperref}
\usepackage[style=apa]{biblatex}

\title{Identifying public values and spatial conflicts in urban planning}

\date{} 

\author[1, 4]{Rico Herzog}
\author[1, 2]{Juliana E. Goncalves}
\author[1, 3]{Geertje Slingerland}
\author[2]{Reinout Kleinhans}
\author[4]{Holger Prang}
\author[1]{Frances Brazier}
\author[1]{Trivik Verma}

\affil[1]{Faculty of Technology, Policy and Management, TU Delft, Netherlands}
\affil[2]{Faculty of Architecture and the Built Environment, TU Delft, Netherlands} 
\affil[3]{Kenniscentrum Creating 010, Hogeschool Rotterdam, Netherlands}
\affil[4]{City Science Lab, HafenCity University Hamburg, Germany}

\begin{document}

\maketitle



\textbf{Keywords}: Public Values, Urban Planning, Public Participation, Spatial Conflict, Natural Language Processing

\begin{abstract}
Identifying the diverse and often competing values of citizens, and resolving the consequent public value conflicts, are of significant importance for inclusive and integrated urban development. Scholars have highlighted that relational, value-laden urban space gives rise to many diverse conflicts that vary both spatially and temporally. Although notions of public value conflicts have been conceived in theory, there are very few empirical studies that identify such values and their conflicts in urban space. Building on public value theory and using a case-study mixed-methods approach, this paper proposes a new approach to empirically investigate public value conflicts in urban space. Using unstructured participatory data of 4,528 citizen contributions from a Public Participation Geographic Information Systems in Hamburg, Germany, natural language processing and spatial clustering techniques are used to identify areas of potential value conflicts. Four expert workshops assess and interpret these quantitative findings. Integrating both quantitative and qualitative results, 19 general public values and a total of 9 archetypical conflicts are identified. On the basis of these results, this paper proposes a new conceptual tool of “Public Value Spheres” that extends the theoretical notion of public-value conflicts and helps to further account for the value-laden nature of urban space.
\end{abstract}

\section{Introduction}
With rapid development, growth and migration in cities, public values are becoming more diverse \parencite{McAuliffe2019ThePluralism}. For more development or redressing urban challenges, a key challenge for planners and decision makers is to identify and address these diverse and often competing values of citizens and other stakeholders \parencite{VanderWal2015From2012}. Integrated and participatory urban development, as highlighted by both the UN’s Sustainable Development Goal 11 and the EU’s New Leipzig Charter, contributes to enhancing the common good and to transitioning to a sustainable and inclusive city \parencite{informal_ministerial_meeting_on_urban_matters_new_2020, united_nations_general_assembly_transforming_2015}. However, the identification of the public’s underlying values and their spatial conflicts through participatory approaches is a complex and challenging task \parencite{Nabatchi2012PuttingValues} as it requires methods to account for diverse and pluralistic public values related to development and usage of urban space \parencite{Campbell1996GreenDevelopment, Godschalk2004LandCommunities, Lombard2016}. Viewing urban space as being continuously socially produced and relational \parencite{Lefebvre1991TheSpace, purcell_theorising_2022}, such a conceptualization often lacks practical application in urban planning to bolster government action through participatory means of governance \parencite{Lehtovuori2016ExperienceSpace}. 

Several scholars have discussed the presence of values in the public sphere \parencite{Bozeman2007PublicInterest,graeber2013value, Hillier1999WhatValues}, including developing public values theory for institutions \parencite{Nabatchi2018PublicGovernance}. They particularly note the importance of inclusive elicitation of public values \parencite{Nabatchi2012PuttingValues}. Literature also highlights the usefulness and importance of mapping (assumed) stakeholder values and demonstrates how participatory mapping can support identification of areas with potential land-use conflicts and provide resolution among stakeholders \parencite{Brody2004ConflictTexas, Brown2014MethodsMapping}. While these studies are highly influential, there is scope to improve current approaches in what they lack, in terms of implicitly assuming value(s) or reducing conflict to a pair of values. A pre-specified list of values \parencite{Brown2014MethodsMapping, Karimi2015MethodsHotspots} leads to a mapping of pre-imposed values while empirical identification of values infers values from data provided by citizens in which they express their values without constraints \parencite{Nabatchi2012PuttingValues}. A narrow investigation on either a singular value or on a dichotomic value conflict \parencite{Tyrvainen2007ToolsAreas} does not recognize the need for a pluralistic approach to understanding the inherent conflicts in urban development \parencite{McAuliffe2019ThePluralism}. Thus, an empirical and cohesive approach is necessary to support inclusive urban development.

Inclusion of public values as one “of the most important aspirations for public participation programs” \parencite[p. 588]{Beierle2000ValuesPlanning} lies at the core of integrated urban development. That is often reflected in the daily work of planners \parencite{Forester1999TheProcesses}, but not in their own job perception \parencite{Lehtovuori2016ExperienceSpace}. Following the calls of \textcite{Nabatchi2012PuttingValues} to leverage participatory data for the identification of public values and their conflicts, it becomes apparent that there are only limited attempts to theoretically establish, and empirically identify, value conflicts with the help of large-scale public participatory data in an urban context. Seminal theoretical work focused on analyzing value conflicts in urban regions proposed a sustainability/livability prism to identify and discuss key values in a relational setting, both physically and metaphorically \parencite{Godschalk2004LandCommunities}. These findings implicitly mark values as static. However, this important work can be expanded to a continuously changing landscape of values in urban areas that are also geolocated and manifest various types of dynamic conflicts. Further, in evaluating several methods to identify social-ecological hotspots, \textcite{Karimi2015MethodsHotspots} call for additional case studies to include social values in spatial decision support tools. 

By building upon the theory of public values, this paper proposes a new approach: A case study-mixed methods design aids in understanding which public values are present and how they can lead to conflict in urban space. Participatory data from an open-source digital participation platform (DIPAS) deployed by the state of Hamburg, Germany with on-site and online participation are integrated into a single public participation geographic information system (PPGIS) for urban planning purposes \parencite{Lieven2017DIPASParticipation}. Structural topic modelling (STM) is utilized to identify public values and spatial clustering algorithms identify areas of potential value conflicts. Lastly, expert workshops serve as a means to qualitatively discuss and interpret the findings of the quantitative analysis. This approach enables inductive identification of public values based on large-scale participatory data \parencite{Thoneick2021IntegratingSystem} and does not require a prior list of values for mapping. It opens up the avenue to extend the present theories on values in urban planning by empirical findings which we synthesize into the conceptual model of public value spheres in urban planning. 

Our goal is two-fold: First, to illustrate the potential of open-source participatory data to elicit geolocated public values in an urban region and identify potential conflicts associated with urban development projects. Second, to advance the present theory in urban studies by translating the empirical evidence to a conceptual model of public values in urban planning that supports in understanding their pluralism across time and space. The paper is structured as follows: First, we lay out seminal theoretical work on public values and illustrate the importance of identification of values for urban planning. Second, we explain the data and methods used. Third, we state the quantitative and qualitative results separately and then integrate the findings of both strands to form a more comprehensive model of conflicting public values in urban planning. Lastly, we discuss our approach and state conclusions for the application of the conceptual model and future research.

\section{Public Values and Urban Planning}
\subsection{Identification of Public Values}
Several scholars in various fields - Anthropology, Philosophy, Psychology, Economics or Public Administration research - have developed a distinct theory of value(s) for their purposes \parencite[e.g.][]{Bozeman2007PublicInterest, Graeber2001TowardDreams, Hillier1999WhatValues}. The most profound and reoccurring distinction is the one between singular and plural: A value (singular) is typically conceived as something tangible, allocatable and traceable in relation to a specific object, for instance the monetary exchange value of a property \parencite[]{graeber2013value}. Values comprise “both cognitive and emotive elements”, are connected to ones definition of oneself, are hardly changeable, and have “the potential to elicit action” \parencite[p. 117]{Bozeman2007PublicInterest}. \textcite{Hillier1999WhatValues} further differentiates between instrumental and intrinsic values, where the former are a means to an end and the latter values are an end in themselves. Distinguishing between an individual’s values for their personal life and the values that ought to be present in the public sphere, the theory of public values is concerned with the values that are followed and embraced by public institutions \parencite[]{Bozeman2007PublicInterest}.

The theory of public values is concerned with “those providing normative consensus about [...] the principles on which governments and policies should be based” \parencite[p. 13]{Bozeman2007PublicInterest} and “the social standards, principles, and ideals to be pursued and upheld by government agents and organizations” \parencite[p. 60]{Nabatchi2018PublicGovernance}. Although the literature recognizes that in reality, the utopia of normative consensus on values seldom materializes \parencite[]{Nabatchi2012PuttingValues}, pluralism of equally valid, correct and fundamental values exists. The “most fundamental” \parencite[]{Fukumoto2019PublicMissing} challenge in this research area is the identification of public values. Multiple approaches have been proposed to this purpose, including analysis of governmental documents \parencite[]{Fukumoto2019PublicMissing}, intuition, elections, surveys and academic literature \parencite[pp. 133-141]{Bozeman2007PublicInterest}. \textcite{Nabatchi2012PuttingValues}, however, argues that any approach for public values identification other than including the public, tends to be exclusionary by highlighting only certain privileged values.

\subsection{Public Values and Urban Space}
The role of values in urban planning gained significant traction since the recognition of relational urban space and its inherently social production \parencite[]{jacobs2016death, Lefebvre1991TheSpace, Soja2000ThirdspacePlaces}. For instance, scholars have described planners as “practical ethicists” \parencite[p. 31]{Forester1999TheProcesses}, rejected the idea of value-neutral planning \parencite[]{Sandercock2004TowardsCentury}, and urged planners to reconsider the plurality of values in play and explore more effective value-incorporation strategies for inclusive urban development \parencite[]{Hillier1999WhatValues}. \textcite{Godschalk2004LandCommunities}, building on the work of \textcite{Campbell1996GreenDevelopment}, partially addressed this issue by discussing values inherent to the planning paradigms of “Smart Growth” and “New Urbanism”. They identified and described four main values, namely economic development, ecologic quality, social equity and livability, that are illustrated as vertices of a prism. More recently in Urban Studies, several articles investigated in how urban values relate to urban qualities \parencite[]{metzger_contested_2018, molnar_framing_2022}.

\subsection{Identification of Value Conflicts in Space}
The existence of value conflicts in urban space is uncontested \parencite[]{deGraaf2016CopingConflicts, Hillier1999WhatValues, Nabatchi2018PublicGovernance}. Scholars point out that “emotional” aspects towards land and places are an integral part of urban life and its resulting conflicts in cities across the globe \parencite[]{Lombard2016, mcmichael2016}. Essential theoretical work in the field was laid down by \textcite{Campbell1996GreenDevelopment} and \textcite{Godschalk2004LandCommunities} by outlining the value conflicts inherent to the sustainable and livable development of cities. \textcite[p. 298]{Campbell1996GreenDevelopment} identified a “triangle of conflicting goals” between environmental protection, economic growth and social equity goals: There is a property conflict between economy and equity, a development conflict between equity and the environment and a resource conflict between economy and ecology. Later, \textcite{Godschalk2004LandCommunities} expanded on these conflicts by adding a fourth dimension, namely “livability”, as an additional public value in urban planning. They identified a gentrification conflict between equity and livability, a green cities conflict between livability and ecology and a growth management conflict between economic and livability values. They incorporated all these conflicts in a sustainability/livability prism investigating how the “ecology of plans” in a city addressed and resolved each of the conflicts; hence assuming that all of these conflicts are already present at a large scale and static. 
In a democratic development of urban space, there is a continuous production of conflicting values \parencite[]{purcell_theorising_2022}, which, by their very nature, are dependent on the person expressing them, and thus incommensurable between people \parencite[]{Bozeman2007PublicInterest, graeber2013value, Rokeach1973}. Value conflicts materialize in space where the realization of one value oftentimes prohibits or impairs the realization of another. From a relational viewpoint, value conflicts are not only unavoidable, but inherent to the planning process itself and are continually changing. 
Even though multiple scholars provide insight into how analyzing the framing elicits urban values \parencite[]{metzger_contested_2018, molnar_framing_2022} and others advocate for increased deliberation and participation in urban planning \parencite[]{Nabatchi2012PuttingValues}, a coherent overview of public values in urban planning is yet to be obtained \parencite[]{McAuliffe2019ThePluralism}. The question that is then imposed on urban decision makers is how to transcend the incommensurability of such barriers and identify public values and their conflicts in urban planning in an empirical and inclusive manner.

\section{Methods and Data}
We conducted a case study-mixed methods approach that embeds a mixed methods study within an overarching case study \parencite[]{Guetterman2018TwoReview}. For the mixed-methods design, we followed an explanatory sequential approach with (1) a quantitative research strand followed by (2) a smaller qualitative study \parencite[]{Creswell2010ChoosingDesign}. Finally, the results from both strands (3) were integrated - or “mixed” – and interpreted to form a better conceptual model of (conflicting) public values in urban space. In public values’ elicitation and their conflicts in urban space, the quality of the quantitative analysis of participatory textual data is dependent on the researchers conducting such a study \parencite[]{Chang2009ReadingModels}. To address this concern our approach encapsulates a qualitative strand that makes use of expert knowledge for the purpose of interpretation and evaluation.
We selected Hamburg, Germany, as a case study for three reasons. First, the city faces substantial urbanization challenges with a potential total population increase of 200,000 people by 2040, thereby crossing two million inhabitants \parencite[]{StatistischesAmtfurHamburgundSchleswig-Holstein2019BevolkerungsprognoseFort}. Second, Hamburg’s Senate founded the “Stadtwerkstatt” (engl. “urban workshop”) to stimulate a new planning culture by proactively involving the citizenry in informal participation processes in urban planning \parencite[]{BurgerschaftderFreienundHansestadtHamburg2012MitteilungStadtwerkstatt} and deploys the digital on-site and online participation platform DIPAS on a city-wide scale \parencite[]{Lieven2017DIPASParticipation}. Third, Hamburg’s own transparency law prescribes the publication of information processed within the city’s administration in a central online repository, that is accessible in an anonymous way without any associated costs \parencite[]{Murjahn2016OpenEffects}. 
In the quantitative strand, we leveraged 4,528 citizen contributions from a total of 25 participation processes on the DIPAS platform where people input open textual comments on local urban planning matters. Of these contributions, 3,584 were geolocated. Preprocessing of the data included basic data cleaning, named entity removal, lemmatization, n-gram modelling and stopword, number and punctuation removal \parencite[]{Grimmer2013TextTexts}, described in more detail in the supplementary information. Subsequently, the data was transformed to a bag of words and the most infrequent 0.75\% of words were removed before applying Structural Topic Modelling (STM) \parencite[]{Roberts2019Stm:Models}, an unsupervised clustering algorithm for large text corpora to infer latent topics behind documents in the text corpus while allowing metadata to influence topic assignments. Additionally, using the Google Translate API, the contributions were translated to English and sentiment analysis was applied to add more nuanced metadata to STM. We chose k = 30 topics as a hyperparameter based on the metrics of residual analysis and semantic coherence and manually assigned topics to a public value, wherever such an assignment was justifiable (see supplementary information). To investigate areas of potential public value conflicts, the different assigned public values were spatially clustered using the HDBSCAN algorithm \parencite[]{Campello2015HierarchicalDetection} and converted into spatial polygons by applying the alpha shape algorithm, as illustrated by \textcite[]{Chen2019UnderstandingInformation}. Using this method, areas of potential public value conflicts can be investigated by exploring the intersection of polygons. 
In the qualitative strand, four expert workshops were conducted with four urban planners based in Hamburg specialized in different areas of urban planning: green space and playground, residential development, mobility, and noise protection. Workshops as a research methodology are designed to serve the professional interests of workshop participants as well as to “produce reliable and valid data about the domain in question” \parencite[p. 72]{rngreen2017WorkshopsMethodology}. Due to restrictions regarding the spread of the Coronavirus and time constraints, workshops were conducted individually in an online Zoom environment. The Stadtwerkstatt Hamburg acted as an intermediary to establish contacts with planning experts. To ensure a mutual benefit of the workshops, only planning experts with background knowledge in the DIPAS platform and participation processes were selected. Expert workshops were structured in three main parts: Section one, “Introduction”, set the scene and introduced participants to the concept of public values. Section two, a short semi-structured interview, sought to gain insights into the perception of public values and their conflicts by expert planners by asking specific questions (see supplementary information). Section three, an interactive discussion, brought the workshop character into play: Together, the STM results were investigated and via remote controlled screen sharing, an interactive map of contributions and areas of possible public value conflicts were explored. Afterwards, workshops were both transcribed and coded for the extraction of relevant knowledge. 
Lastly, both quantitative and qualitative research strands were integrated  to formulate a more comprehensive model of public values and their spatial conflicts in urban planning. By building on existent metaphorical descriptions of pluralistic values as spheres and universes \parencite[]{graeber2013value, VanderWal2015From2012} we suggest different visualizations that depict the identified public values and spatial conflicts as a snapshot in the case study of Hamburg.

\section{Quantitative Results}
\begin{figure}
     \centering
     \begin{subfigure}[b]{0.45\textwidth}
         \centering
         \includegraphics[width=\textwidth]{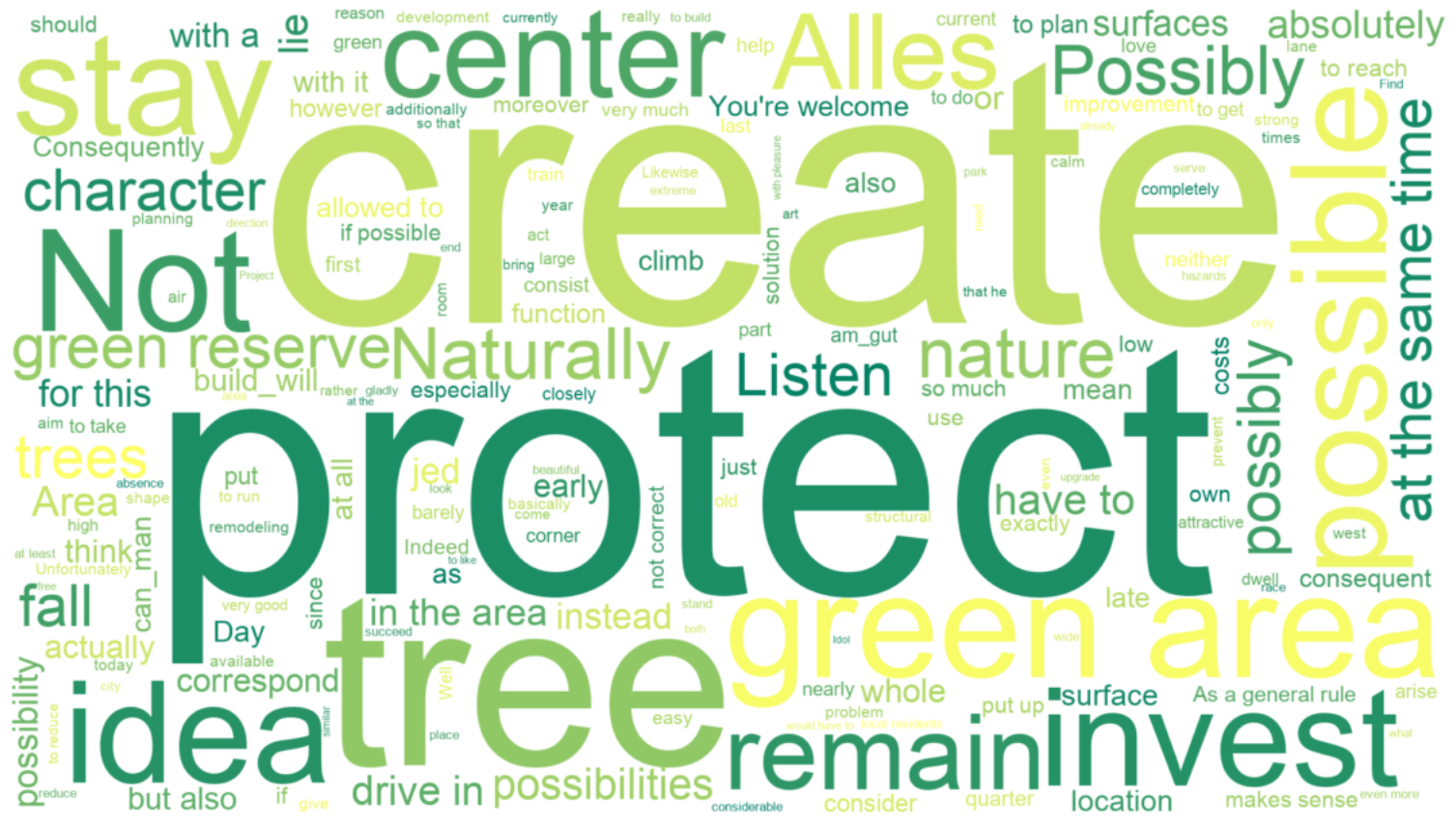}
         \caption{The public value of ecologic quality centres around the creation and protection of green spaces.}
         \label{fig:wordcloud_ecology}
     \end{subfigure}
     \hfill
     \begin{subfigure}[b]{0.45\textwidth}
         \centering
         \includegraphics[width=\textwidth]{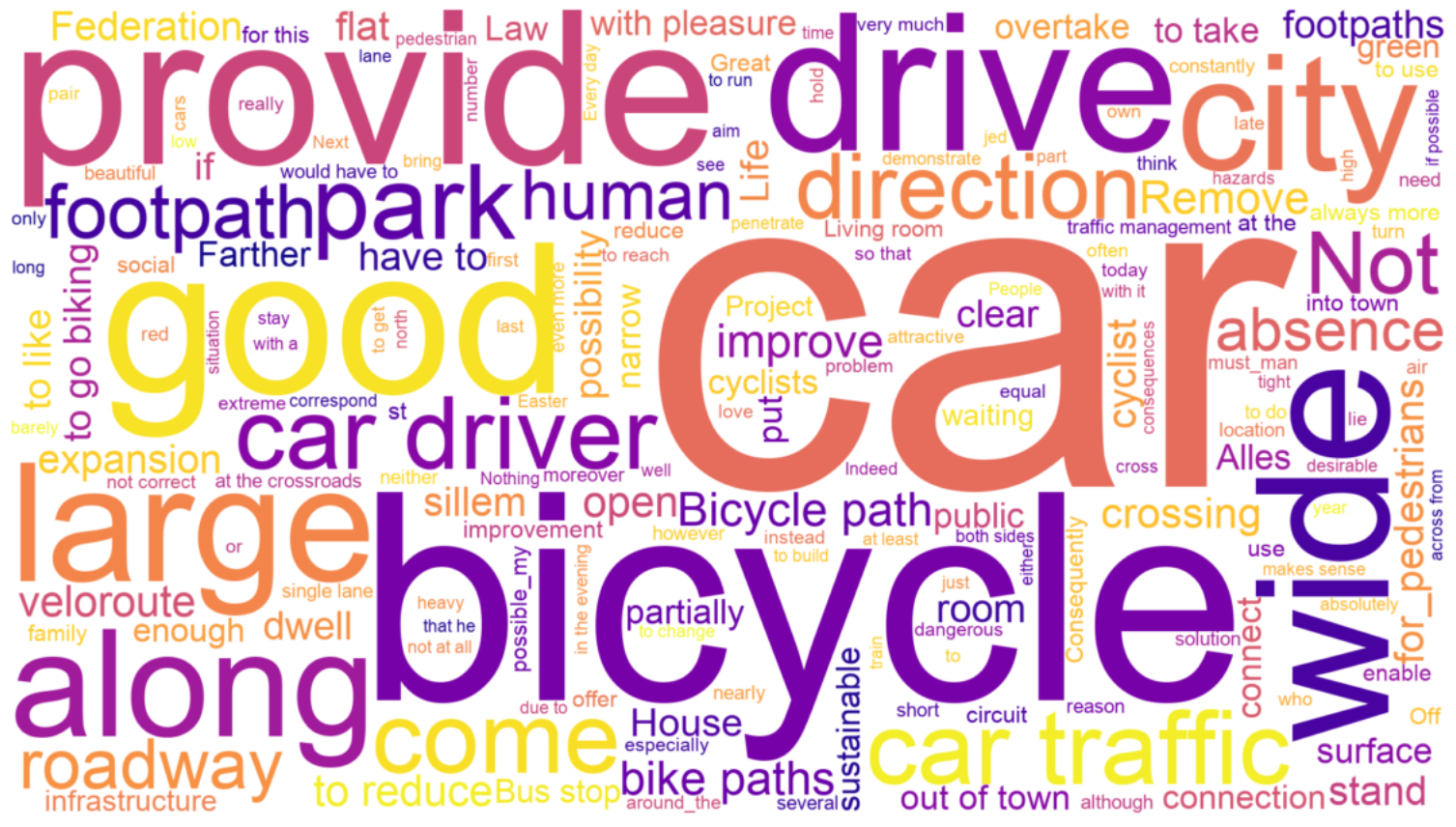}
         \caption{The public value of social equity mainly reflects the provision of better bicycle lanes for better accessibility.}
         \label{fig:wordcloud_social_equity}
     \end{subfigure}
     \hfill
     \begin{subfigure}[b]{0.45\textwidth}
         \centering
         \includegraphics[width=\textwidth]{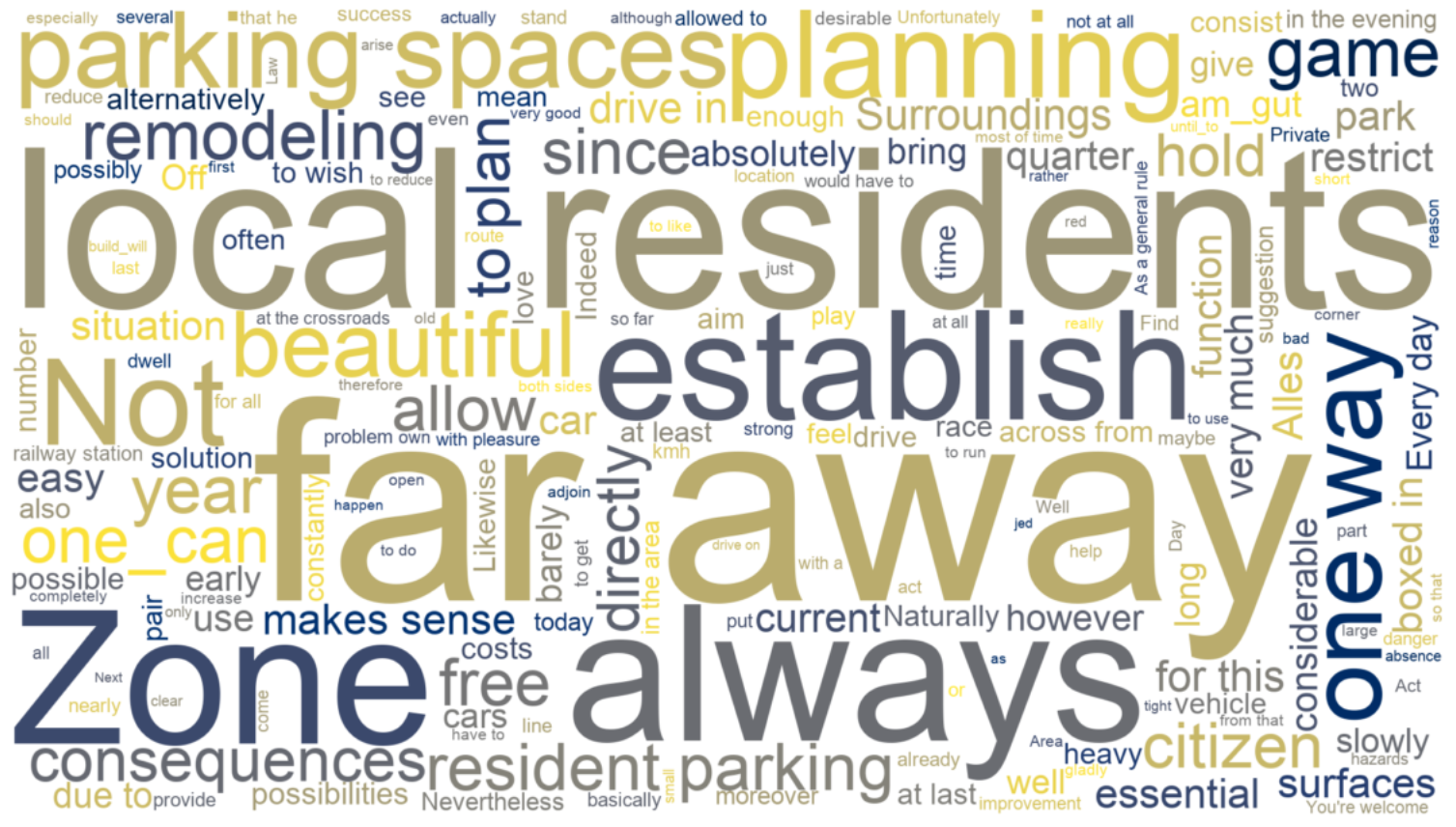}
         \caption{The public value of economic opportunity is largely concerned with the provision of parking spots for local residents.}
         \label{fig:wordcloud_economic_opportunity}
     \end{subfigure}
     \hfill
     \begin{subfigure}[b]{0.45\textwidth}
         \centering
         \includegraphics[width=\textwidth]{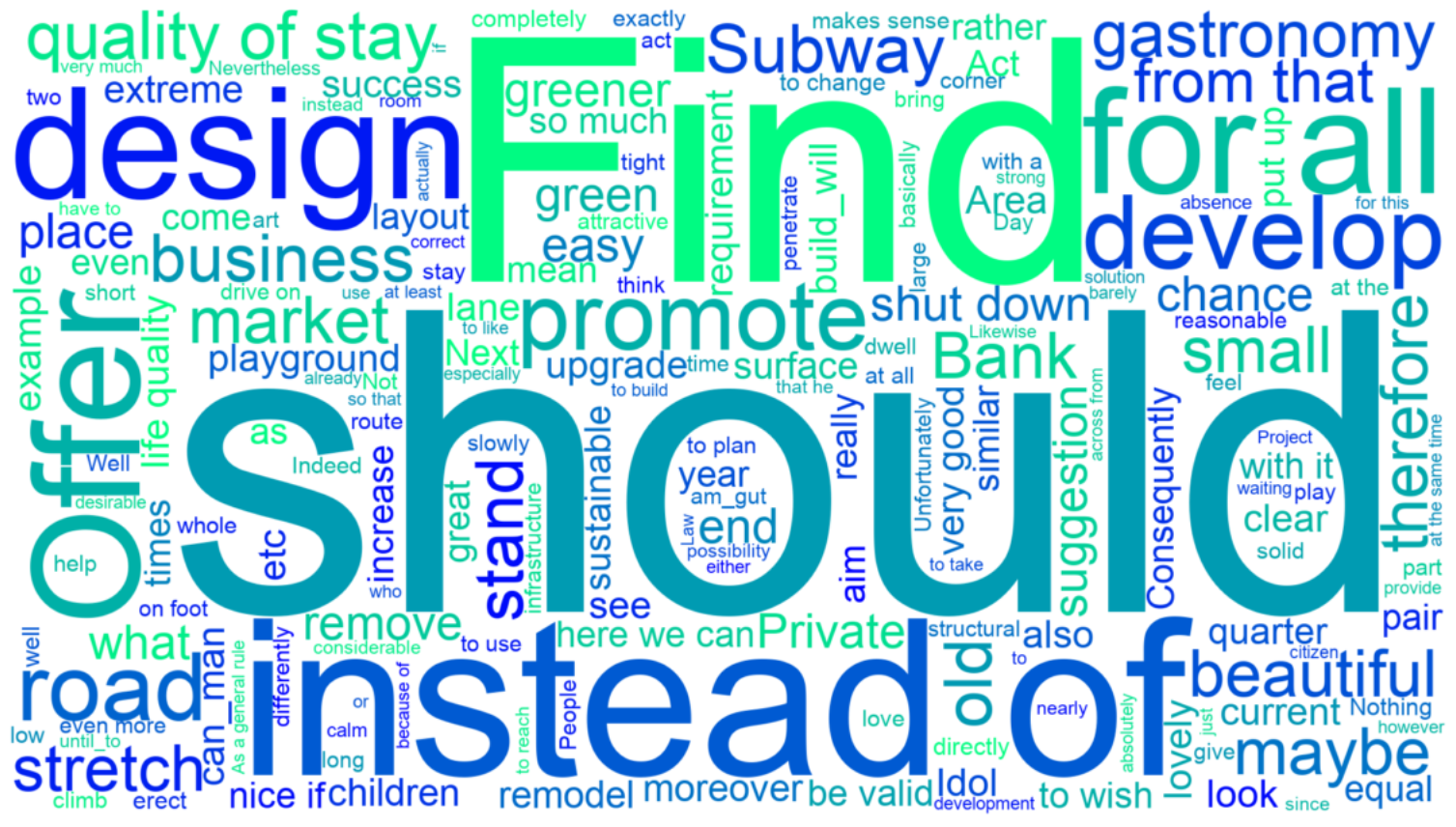}
         \caption{The public value of livability reflects certain wishes for the built environment regarding design that should be realized.}
         \label{fig:wordcloud_livability}
     \end{subfigure}
     \hfill
     \begin{subfigure}[b]{0.45\textwidth}
         \centering
         \includegraphics[width=\textwidth]{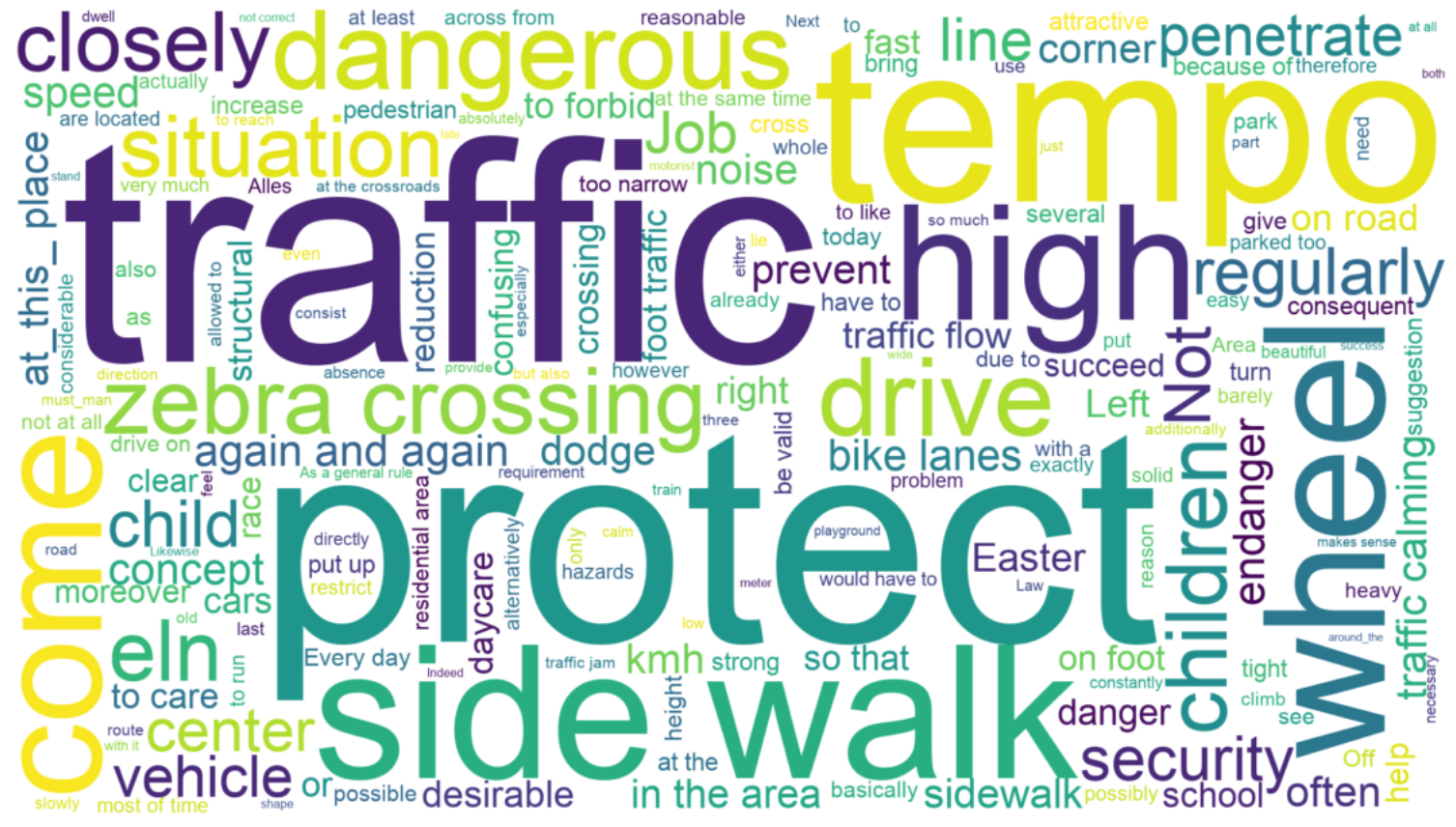}
         \caption{The public value of health/safety indicates the wish for better protection from dangerous traffic situations.}
         \label{fig:wordcloud_health_safety}
     \end{subfigure}
     
    \caption{Wordclouds show the most probable words that appear in topics assigned to a single public value. }
    \label{fig:wordclouds}
\end{figure}

Using STM on the participatory data of citizens in Hamburg, 30 topics were identified that were discussed in relation to urban development projects. 28 topics were assigned with a caption that represents the overarching concepts discussed in each topic, and two showed no broader coherent theme. In these topics, five broad public values were identified: Economic opportunity, ecologic quality, social equity, livability and safety/health. A single public value is reflected in 19 topics; the remaining ones contain various public values (for more details see supplementary information). As shown in Figure \ref{fig:wordcloud_ecology}, the public value of ecologic quality revolves around the two main ideas of protection and creation of green spaces. The most probable words of the public value of social equity as shown in Figure \ref{fig:wordcloud_social_equity} are “car”, “bicycle”, “provide”, “good” and “wide”. They reflect the idea that the city should be more accessible to cyclists by the improvement of bike lanes, e.g. their widening. Contributions of topic 11, “Living for marginalized groups”, in another notion of social equity, lament exploding apartment rents, wish for subsidized and accessible living space for disabled people, suggest increased investment in neighborhood activities due to increasing loneliness of the elderly/the youth or actively look out for partners to provide living space for the homeless. As \textcite{Campbell1996GreenDevelopment} describes the value of economic opportunity realized as the “space of highways, market areas and commuter zones”, the issue of providing parking spots for residents in public spaces was assigned to the public value of economic opportunity. As many contributions are attributed to this specific issue, the most probable words are “local residents”, “far away”, “zone”, “always” and “establish” (see Figure \ref{fig:wordcloud_economic_opportunity}). Representing the more tangible everyday environment and its design, many (partially contradicting) desires are subsumed under livability. Dominant words are “should”, “find”, “instead of” and “design”, indicating a wish for a different shaping of urban space. Looking for terms that concretize these desires, words like “gastronomy”, “subway”, “business”, “market”, “bank” and “playground” can be found. Lastly, the public value of health/safety was identified mainly in mobility-related topics, which explains their most probable words: “traffic”, “protection”, “tempo”, “high”, “sidewalks” and “dangerous situations” (see Figure \ref{fig:wordcloud_health_safety}).

\begin{figure}

    \centering
     \begin{subfigure}[b]{\textwidth}
         \centering
         \includegraphics[width=\textwidth]{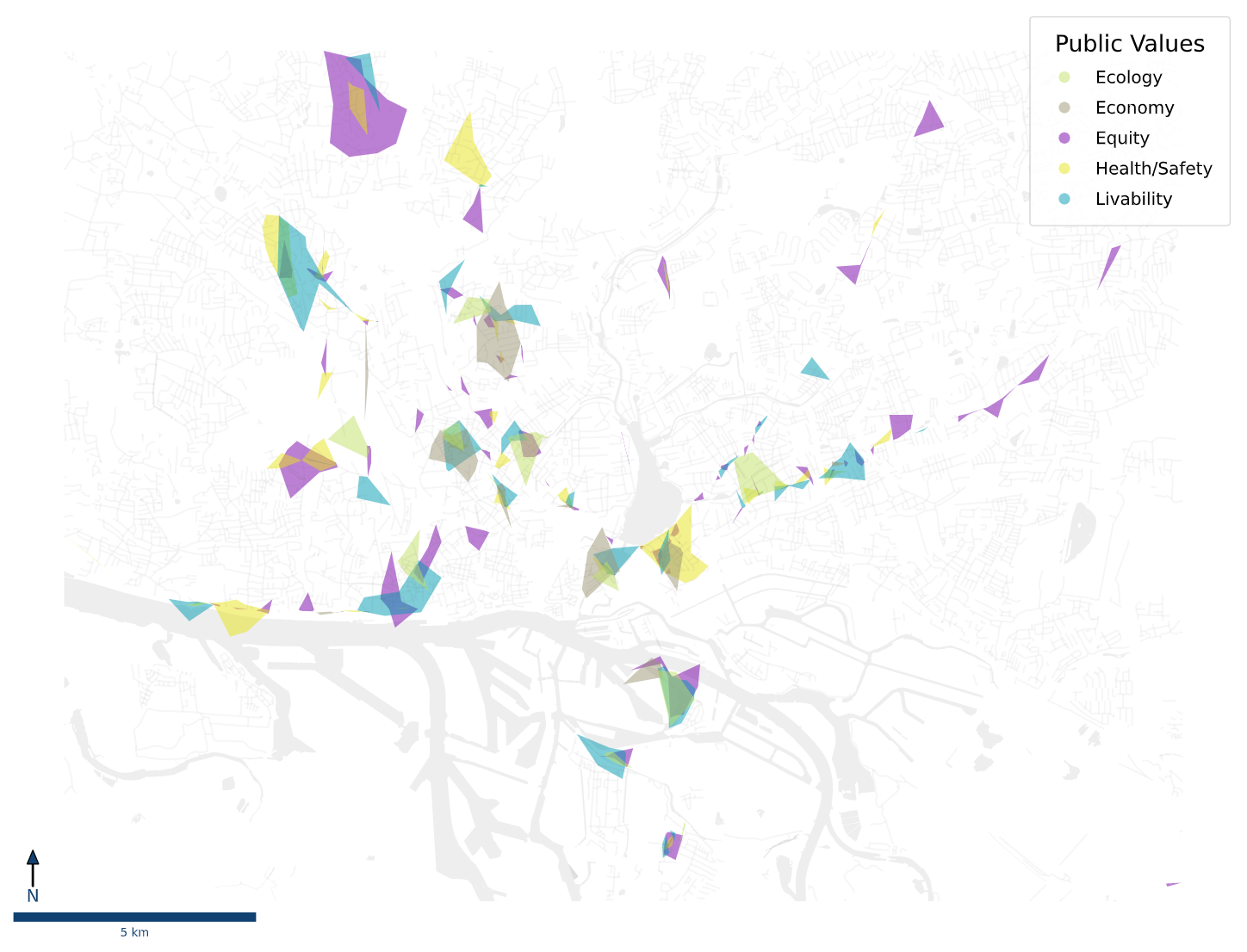}
         \caption{Public Values are distributed unevenly across the city, showing both larger and smaller areas of public value clusters. }
        \label{fig:value_map}
     \end{subfigure}
     \hfill
     \begin{subfigure}[b]{\textwidth}
         \centering
         \includegraphics[width=\textwidth]{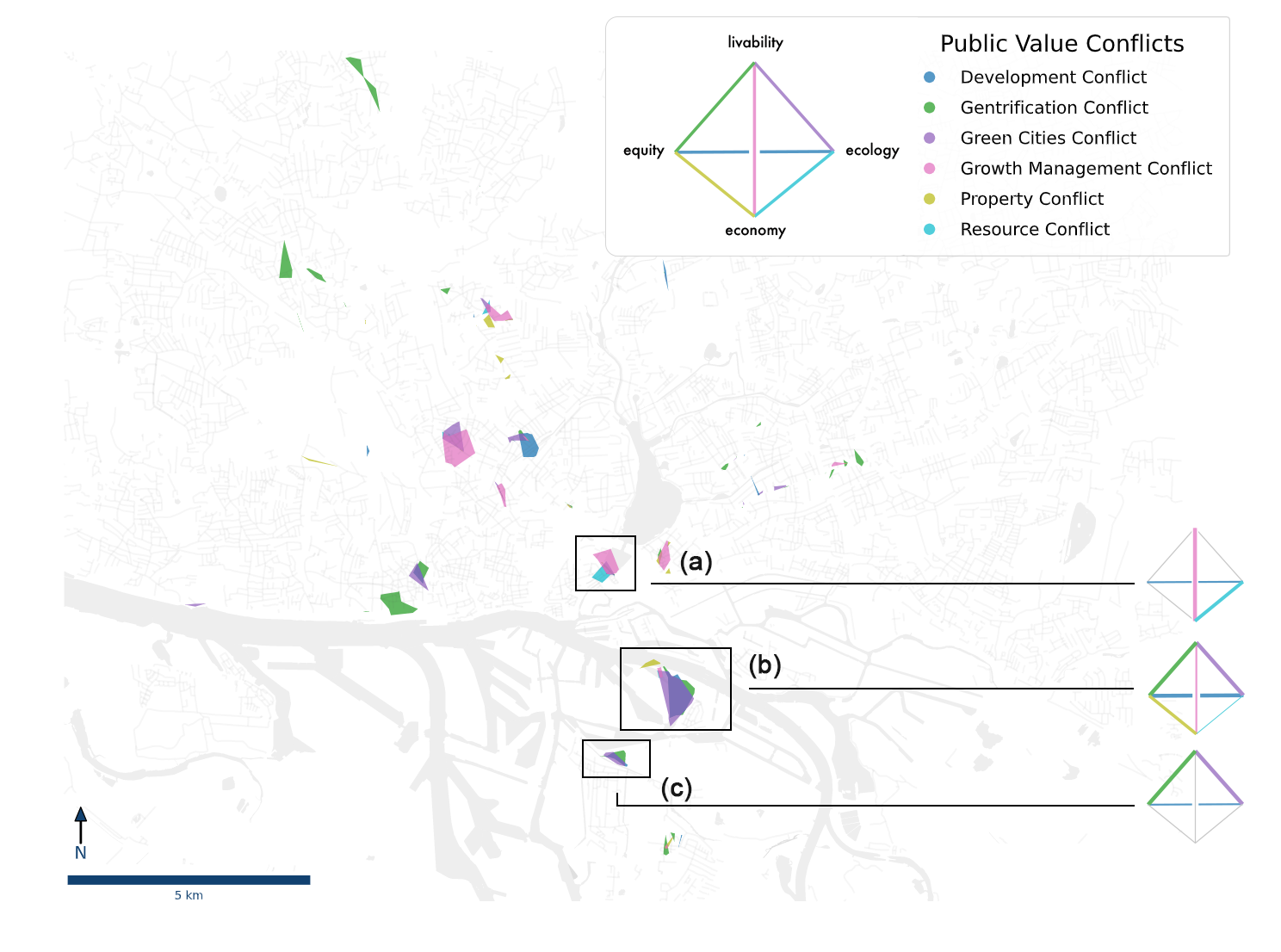}
         \caption{Public Value Conflicts under the sustainability/livability prism appear in multiple parts of the city. Interesting spaces for further insights were selected at the Jungfernstieg [a], the Grasbrook [b] and the Spreehafenviertel [c] sites.}
         \label{fig:conflict_map}
     \end{subfigure}
    \caption{Public Values and their Spatial Conflicts, distributed over the map of Hamburg, Germany.}
    \label{fig:value_conflicts}
    
\end{figure}

Figure \ref{fig:value_map} illustrates the spatial distribution of public value clusters detected by the HDBSCAN and alpha shape algorithm. Note the several larger value clusters of equity, livability and health/safety in this map. Additionally, three larger clusters of economic values can also be observed. Multiple smaller clusters of each value are distributed across the city. Intersections of overlapping clusters show areas of potential public value conflicts (Figure \ref{fig:conflict_map}). Using Godschalk’s prism, six types of conflict are distinguished and illustrated with the help of three regions of special interest in Hamburg \parencite[]{Godschalk2004LandCommunities}.
One, in multiple areas a development conflict manifests. Its most common form is the dedication of street space for increased access to pedestrians and cyclists as opposed to green area protection, specifically along the Jungfernstieg (Figure \ref{fig:conflict_map} [a]). 
Two, in the newly developed district of the Grasbrook (Figure \ref{fig:conflict_map} [b]) and the Spreehafenviertel (Figure \ref{fig:conflict_map} [c]), the gentrification conflict becomes visible in a clash between the wish for both highly livable environments and affordability. As these neighborhoods are about to be developed for residential living, multiple people wish for affordable government housing while other people wish for exciting architecture and other appealing places. 
Three, the green cities conflict is only partially reflected in the case study of Hamburg. In the perception of most contributors, both livable and green spaces go hand in hand. This becomes especially apparent around the Jungfernstieg (Figure \ref{fig:conflict_map} [a]): Frequently, livability values reflect a desire for more urban green instead of the built environment. However, there are also sporadic contribution who wish for open-air cinemas, more art and more kiosks, which could be in a green cities conflict with contributors wishing for the restoration of nature. 
Four, the growth management conflict again becomes visible around the Jungfernstieg [a] area, which was chosen to be car-free as a pilot project. Here, multiple people lament the decay of the inner city due to lacking accessibility of well-funded customers by car. Simultaneously, many see a more livable environment created through the exclusion of private cars.
Five, multiple smaller sections in Hamburg exhibit the property conflict between equity and economic values. The “social character of land” which is in conflict with “its private ownership and control” (Campbell, 1996) is frequently reflected in the wish of redeveloping street/parking space for cyclists and pedestrian. In that way, pedestrian and bike lanes reflect a much more social sharing of space. Private cars are considered to take up public space that could be used otherwise for a broader benefit to society. 
Six, the resource conflict between economic and ecologic values manifests only in very sparse sections of the city. It is identified in similar spaces as the property conflict, since oftentimes, the wish for more parking spots contradicts the wish for the preservation or creation of green areas.

\section{Qualitative Results}
Interviewed about the presence of values and value conflicts, the answers of expert planners depend on their area of expertise. Typically, DIPAS online participation serves a supplementary function to other more formal and direct procedures. Although all planners mention that they read the entirety of contributions, some of them focus on public hearings and actual conversations. As one planner puts it, \textit{\say{We don’t just use the online tool. This online participation is always supplemental. We always focus on personal encounters, personal information and personal exchange. Because such conflicts can be discussed and weighed up in a completely different way}} (Workshop participant 3). Additionally, planners draw from formalized participation procedures and involvement of multiple institutions, such as the fire department, official clubs or other nongovernmental organizations. This broader input of participatory data widens the space for public value identification for them.
The value of ecologic quality is identified by all of the workshop participants. According to them, public spaces should not be sealed and present green areas shall be preserved: \textit{\say{There’s something that would be widely perceived as a forest, and citizens who were consulted want to preserve this as far as possible} }(Workshop participant 3). Citizens wish for the planting of (fruit) trees, the creation of space for bees and insects and the conversion of sealed parking space to green space. 
The value of social equity revolves around inclusion, the creation of spaces for everyone and accessibility both as a means to enable a better access of citizens with disabilities and as a goal to increase the accessibility of other modes of transportation than the car. As workshop participant two put it, \textit{\say{[...] there are many people, especially in inner-city areas, who consciously decide not to own a car and then say they would rather use the space for something else.}}
The value of economic opportunity was most apparent in a noise protection context: \textit{\say{There is still little that can be done about aircraft noise because the economic component is so strong. We can’t do anything about it, we can’t do anything at all.}} (Participant Workshop 4). In other instances, residential development was mentioned: \textit{\say{People say \say{Yes, we want this progress. We want housing and we want something to develop and happen here}}} (Workshop participant 3).
The value of livability has many different facets; it is an umbrella term that bundles a myriad of individual perceptions: e.g. the public values of sports, aesthetics, cleanliness and quietness. Valuing sports - in the perception of one planner - it is a temporary fashion: \textit{\say{That’s a trend that’s just coming back, so sports in public spaces, accessible to everyone. [...] Sports clubs with fixed times are on the decline and people want to do individual sports. They want more and more sports in public spaces.}} (Workshop participant 1). Citizens also value (historical) aesthetic aspects in their everyday built environment: \textit{\say{I would say there is a nostalgia factor. You live in […] a cobblestone street, and you think that’s nice. Of course, that also fits in with the old buildings that are there. Needless to say, it’s worth preserving in certain areas, and it also somehow gives the street its flair and charm.}} (Workshop participant 2). Additionally, people value cleanliness in their surroundings: \textit{\say{Quite often when we have downtown playgrounds, people say \say{Build a public restroom!}} }(Workshop participant 1). Quietness is another value that appears regularly in the context of the everyday lived environment: \\textit{say{So we also notice that with certain age groups it’s quite normal, if they just meet in the group and talk, then it gets louder. They don’t even shout, but they are always perceived as a point of disturbance} }(Workshop participant 1).
The public value of safety is primarily related to traffic. In this specific domain, \textit{\say{Safety is always the top priority and I have to subordinate everything to it.}} (Workshop participant 2). More specifically, \textit{\say{it’s [...] about individual safety and also about the perception of safety. So it’s not always about objective safety. Rather, what plays a role is subjective safety.}} (Workshop participant 2). In that conceptualization of safety as the absence of fear, it is to be distinguished from the public value of health. Although analyzed together in the quantitative strand, discussions with planners reveal that health in the public perception might be decoupled from safety. As one planner puts it, \textit{\say{[...] otherwise, things like fine dust pollution or something like that would certainly play a role.}} (Workshop participant 2), also linking the topic of health with the topic of ecologic quality. Additionally, health is also related to the public value of quietness: When talking about a noise level map of Hamburg, one planner admits: \textit{\say{If you look at the map as an overview of Hamburg, everything is red or everything is purple. There is a lot that is already a health hazard.}} (Workshop participant 4). Lastly, multiple planners refer to people opposing action or change. One planner mentioned \textit{\say{that ‘everything was better in the past’ comes up quite often.}} (Workshop participant 1). Another expert, when asked for which public values citizens would like to see realized, answered: \textit{\say{I would say there are not so many things that they wanted to see realized as things that they did not want to see realized.}} (Workshop participant 3). This value might well be titled conservatism in a sense of conserving the status quo. 
Expert workshops brought forward a wide range of public value conflicts. Most of the conflicts outlined by \textcite{Godschalk2004LandCommunities} manifest in the case study of Hamburg. Additionally, we identify three value conflicts which we call the \say{Dangers of Nature Conflict}, the \say{Externality Conflict} and the \say{Drawback of Beauty Conflict}. For the \say{Dangers of Nature Conflict}, a planner describes how the value of safety opposes ecological values: \textit{\say{So when I think about ecology and fruit trees and bees, they all yell for it. But if I plant fruit trees in a playground and in the summer the bees come, the parents don’t like it either.} }(Workshop participant 1). The \say{Externality Conflict} appears in between economic development viewpoints and the public value of health when the externalities of economic development (such as noise and/or pollution) impair the health of residents. The \say{Drawback of Beauty Conflict} between the values of tranquility and aesthetics manifests in ways that aesthetic spaces typically attract people, which will then lead to noise.

\section{Integration}
The results of both strands showcase that there is high agreement of public values and conflicts that were identified. In the qualitative strand, the findings of the quantitative strand were largely confirmed and valuable and contextual information from other sources of participatory data were gathered. For instance, the public value of conservatism and several public values inside the umbrella term of livability were mentioned and three additional conflicts were identified. However, integrating both strands, it is of dialectic interest to look specifically into points of discrepancy. The resource conflicts in this regard provides an exemplary instance: Although planners describe the conflict between nature preservation and economic development as an archetypal conflict in every new development project, the quantitative results barely reflect that. This is not due to lacking citizen contributions which can be frequently found on both sides, but rather due to improper assignments and hence missing spatial clusters. Additionally, value conflicts that open up between official municipal decisions, institutionalized actors and citizens are not identifiable in the quantitative strand. In that way, the qualitative results diverge from the quantitative due to the expert planner’s access to different kinds of participatory data.
Overall, the integration of the quantitative and qualitative strands demonstrates that public values and their conflicts in urban planning are not limited to the ones outlined by \textcite{Godschalk2004LandCommunities}. The sustainability/livability prism can be found to serve as a solid conceptual tool for identifying public value conflicts inherent to sustainability and livability. However, these public values and conflicts only cover a certain part of the total spectrum of public values in urban planning; hence creating the need to expand the prism towards a more comprehensive and encompassing conception of public values.

\begin{figure}
    \centering
    \begin{subfigure}[b]{0.8\textwidth}
         \centering
         \includegraphics[width = \textwidth]{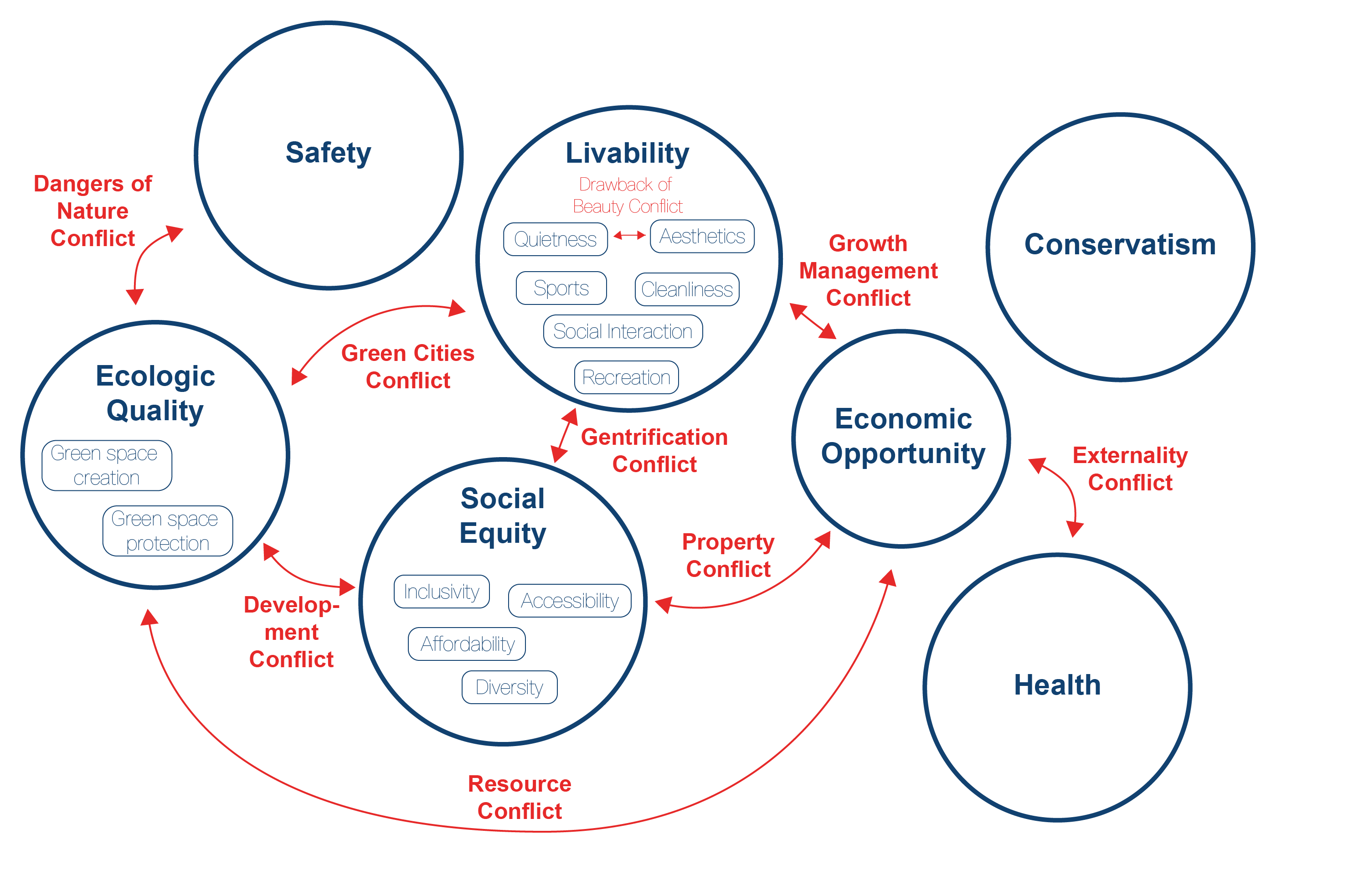}
        \caption{Public Value Spheres outline different archetypal conflicts manifesting between values in urban space.}
        \label{fig:value_spheres_conflicts}
     \end{subfigure}
     \hfill
     \begin{subfigure}[b]{0.8\textwidth}
         \centering
         \includegraphics[width = \textwidth]{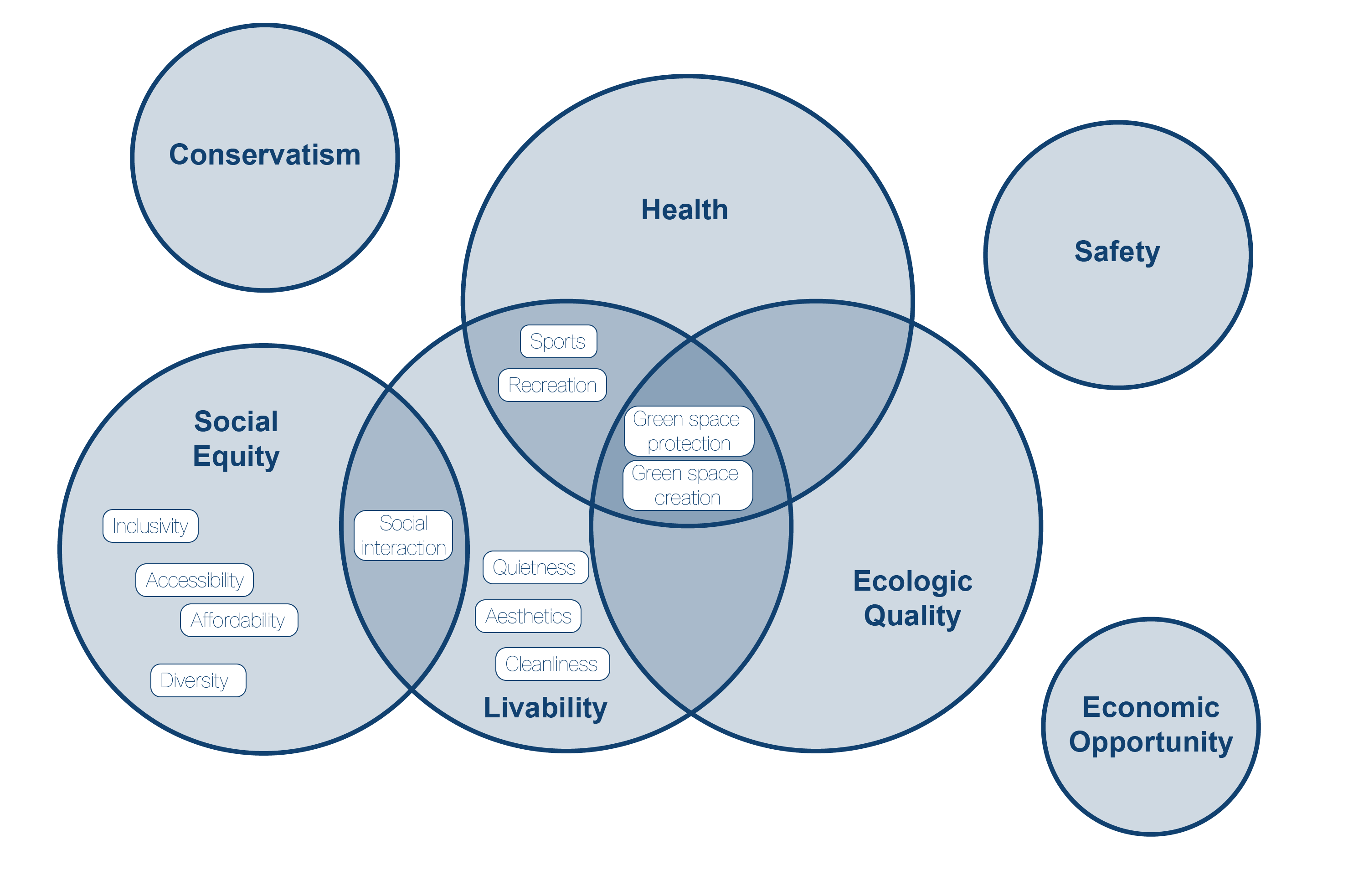}
        \caption{Public Value Spheres display the possible connections between instrumental and intrinsic values.}
        \label{fig:value_spheres_overlaps}
     \end{subfigure}
    \caption{Different arrangements of public value spheres allow for a closer investigation in relational urban space}
    \label{fig:value_spheres}
\end{figure}

Similar to exemplary scholarly work by \textcite{Jrgensen2007PublicInventory, VanderWal2015From2012, graeber2013value} who make use of the metaphor of universes, galaxies and value spheres when conceptualizing the plurality of (public) values, this paper proposes to expand the sustainability/livability prism of \textcite{Godschalk2004LandCommunities} to public value spheres in urban planning. Leaving behind the enclosed volume of a prism, conceptualizing public values in spheres models a much more pluralistic view on public values and their conflict in urban space. Contrary to geometrical shapes that grow in complexity when adding vertices, a spherical representation of public values allows for a simple expansion and display of the pluralism of public values and their conflicts across different spatial and temporal scales. Thus, the model allows for different snapshots and arrangements of public values to become an important element to capture, visualize and analyze different (conflicting) public values in participatory settings and beyond. At the same time, the conceptual work of previous authors is embedded in a larger context.
Two exemplary figures showcase Hamburg-specific snapshots of city-scale public values and their conflicts during the time of data gathering from 2018-2021. Based on the integrated findings of both research strands, we display which conflict archetypes we found in between public values in Figure \ref{fig:value_spheres_conflicts} and how public values are interconnected and relate to each other in Figure \ref{fig:value_spheres_overlaps}. In the former depiction, the conflicts previously outlined are mapped in between value spheres. In the latter depiction, the intersections of several public value spheres show where certain instrumental values might be connected to intrinsic values.

\section{Discussion}
This paper illustrates the potential of participatory data in eliciting geolocated public values and presents a new conceptual model of (conflicting) public values in urban planning based on our integrated findings. The novel case-study mixed methods approach provides an innovative and viable way to combine quantitative and qualitative work in an urban context. From a methodological point of view, connecting natural language processing with spatial clustering algorithms adds to the list of spatial analysis techniques, specifically the ones for participatory big data. Mixing methods with qualitative expert workshops succeeded in adding nuance and context, since applying quantitative methods to largely unstructured textual data in many cases can still not outperform human judgement \parencite[]{Chang2009ReadingModels}. This is even more the case when inferring public values from citizen contributions, a task that is very much dependent on background information and knowledge of the human nature \parencite[]{Bozeman2007PublicInterest}. In that regard, the qualitative follow-up mitigates the shortcomings of the quantitative methodology and supplements its advantages in analyzing big participatory data. From a theoretical point of view, this paper contributes by bringing together public value theory originating in public administration research with urban planning and public participation research. Utilizing the concept of relational space as a common theme, the proposed conceptual model of public value spheres provides an extended theoretical understanding on how citizens value urban space and how these valuations might lead to conflicts in urban development, showcased on the case study of Hamburg. 

\subsection{Applying Public Value Spheres}
Specifically, for the case study of Hamburg, this paper identifies 19 main public values and a total of nine archetypal conflicts that are mapped out in two snapshots of public value spheres that show both the present public values and their archetypal conflicts. More broadly, public value spheres provide support for a better understanding of the plurality of values attached to urban space, their interrelations and their archetypal conflicts. Specifically for urban planning and policy, they could be considered to be a conceptual model to map out values and conflicts involved in participatory processes. This may, however, involve rethinking and possibly adapting current (digital) participatory practices to also elicit the underlying public and individual values, i.e. truly understanding \say{why} people want something, not only \say{what} they want. As citizens however might not always be equipped in articulating their underlying values, research on generative design techniques and context-mapping \parencite[]{SleeswijkVisser2005Contextmapping:Practice} might provide additional starting points on how to support the extraction of tacit knowledge and latent needs. Collecting such information could ultimately lead to better decision-making by facilitating the mediation process between conflicting citizen groups. For citizens themselves, public value spheres provide a means to exchange viewpoints and understand how other perspectives relate to the own. They might also serve as a visual means to understand that in a deliberative planning process, there will almost inevitably be conflicts between public values. For advancing scholarly work, public value spheres can provide a means to study and reveal the interconnections of several public values and possibly map them to socioeconomic groups for future sociological and data-driven research. Finally, by taking into account the value-laden nature of urban space and identifying underlying public values, the aim of developing more sustainable and inclusive cities as pursued in the UN’s SDG 11 is supported.

\subsection{Limitations}
Generally, participatory data is subject to bias, especially when there is no random sampling of contributions \parencite[]{Brown2014WhichManagement} and people who are in vulnerable life situations are likely to be underrepresented in the data. As DIPAS was developed with data privacy concerns in mind, neither planners nor platform providers know who contributes to the various processes, thus trading off a possible mobilization of special interest groups against a low-threshold design considering data privacy \parencite[]{Lieven2017DIPASParticipation}. Such biased input data, however, only affects the research to a certain extent since no normative statements are inferred from the public value identified. Additionally, another inherent content bias of the input data is likely: As multiple participation projects were posed with project-specific questions to the citizenry, the topics and public values in the contributions might also reflect these initial conditions. 
From a methodological perspective, the participatory data served solely as a proxy to identify public values. Multiple implicit and explicit assumptions are made when inferring public values from textual contributions, both in the process of manual value assignment, as well as in the process of facilitating expert workshops. Especially the distinction between intrinsic and instrumental values poses a challenge whenever it remains unclear if a certain demand is an end in itself or a means to realize another intrinsic value. 
Please note that the results outlined are embedded in the overarching case study of Hamburg. The geographical and historical context of the city must be acknowledged when analyzing certain specific public values and their conflicts. As the proposed conceptual tool of public value spheres is explicitly designed to be open to expansion and/or rearrangement, future case studies in differing contexts might help with drawing a more holistic picture by providing additional “snapshots” of (conflicting) public values other spatiotemporal settings.

\subsection{Future research}
Multiple alleys for future research open up through this research. Building on the discussion, three main starting points can be explored. First, additional case studies in other social and cultural contexts, especially ones in the Global South, would add additional perspectives on public value spheres. Enlarging this public value network with additional values and archetypal conflicts will ultimately lead to a more comprehensive understanding of the complex interactions of public values and urban space and into how decision makers could possibly mitigate or resolve such conflicts. Second, investigating the differentiation between individual and public values and the sources of conflict in between these values, would certainly help explain multiple urban social phenomena and could provide starting points for more effective public participation processes \parencite[]{Thoneick2021IntegratingSystem}. Third, research into participatory processes and generative design techniques could aid in improving current PPGIS platforms to not only investigate what citizens want, but also their underlying values. 

\section{Conclusion}
Understanding public values and especially their inherent conflicts is crucial to achieve the UN’s Sustainable Development Goal 11 of more inclusive and sustainable cities \parencite[]{united_nations_general_assembly_transforming_2015}. Only once such conflicts are identified, can they be effectively addressed, mitigated and potentially resolved. By viewing such conflicts through the lens of public values spheres, the present case study of Hamburg showcases a new approach to both identify and conceptualize public values in urban space. Attempts to truly understand the citizenry should start at identifying their norms and values; the principles on which the development of their city should be based. Emerging technology and multidisciplinary approaches now enable an integration of pluralistic values obtained from a large-scale sample of citizens for inclusive and sustainable planning of urban space.

\section*{Acknowledgements}
We would like to thank the Stadtwerkstatt Hamburg, especially Mateusz Lendzinski, for the data provision of multiple DIPAS participation processes and for establishing the contacts to the expert planners to conduct the workshops.

\printbibliography

\end{document}